\documentclass[11pt]{article}
\usepackage{graphicx,cite,amsfonts}
\usepackage[left=2cm,right=2cm,bottom=2cm,includeheadfoot,a4paper]{geometry}
\usepackage[onehalfspacing]{setspace}

\newcommand{\T}[2]{T_{#2}^{({#1}d)}}
\newcommand{\Tn}[1]{T_{#1}^{(nd)}}

\begin{document}

\title{Electronic Energy Spectra of Square and Cubic Fibonacci
    Quasicrystals}
\author{SHAHAR EVEN-DAR MANDEL and RON LIFSHITZ\\
  Raymond and Beverly Sackler School of Physics and Astronomy\\
  Tel Aviv University, Tel Aviv 69978, Israel} \date{today}
\date{December 18, 2007}
\maketitle

\begin{abstract}
  Understanding the electronic properties of quasicrystals, in
  particular the dependence of these properties on dimension, is among
  the interesting open problems in the field of quasicrystals. We
  investigate an off-diagonal tight-binding hamiltonian on the
  separable square and cubic Fibonacci quasicrystals. We use the
  well-studied singular-continuous energy spectrum of the
  1-dimensional Fibonacci quasicrystal to obtain exact results
  regarding the transitions between different spectral behaviors of
  the square and cubic quasicrystals. We use analytical results for
  the addition of $1d$ spectra to obtain bounds on the range in which
  the higher-dimensional spectra contain an absolutely continuous
  component. We also perform a direct numerical study of the spectra,
  obtaining good results for the square Fibonacci quasicrystal, and
  rough estimates for the cubic Fibonacci quasicrystal. 
\end{abstract}

\section{Background and Motivation}
\label{sec:intro}

As we celebrate the Silver Jubilee of the 1982 discovery of
quasicrystals~\cite{shechtman}, and highlight the achievements of the
past two and a half decades of research on quasicrystals, we are
reminded that there still remains a disturbing gap in our
understanding of their electronic properties. Among the open questions
is a lack of understanding of the dependence of electronic
properties---such as the nature of electronic wave functions, their
energy spectra, and the nature of electronic transport---on the
dimension of the quasicrystal. In an attempt to bridge some of this
gap, we~\cite{ilan,me} have been studying the spectrum and electronic
wave functions of an off-diagonal tight-binding hamiltonian on the
separable $n$-dimensional Fibonacci quasicrystals\footnote{The reader
  is referred to Refs.~\cite{definition1} and \cite{definition2} for
  precise definitions of the terms `crystal' and
  `quasicrystal'.}~\cite{squarefib}. The advantage of using such
separable models, despite the fact that they do not occur in nature,
is the ability to obtain exact results in one, two, and three
dimensions, and compare them directly to each other. Here we focus on
the energy spectra of the 2-dimensional ($2d$) and 3-dimensional
($3d$) Fibonacci quasicrystals to obtain a quantitative understanding
of the nature of the transitions between different spectral behaviors
in these crystals, as their dimension increases from 1 up to 3. In
particular, we consider the transitions between different regimes in
the spectrum, taking into account the existence of a regime in which
the spectrum contains both singular continuous and absolutely
continuous components. These different behaviors of the
higher-dimensional spectra are expected to reflect on the physical
extent of the electronic wave functions, as well as on the dynamics of
electronic wave packets, and are therefore of great importance in
unraveling the electronic properties of quasicrystals in general.

Recall~\cite{ilan} that the off-diagonal tight-binding model assumes
equal on-site energies (taken to be zero), and hopping that is
restricted along tile edges, with amplitude 1 for long ($L$) edges and
$T$ for short ($S$) edges, where we take $T\geq 1$. The Schr\"odinger
equation for the square Fibonacci quasicrystal in $2d$ (with obvious
extensions to higher dimensions) is then given by
\begin{equation}
  \label{eq:twoDeq}
  T_{n+1} \Psi(n+1,m) + T_n \Psi(n-1,m)
  + T_{m+1} \Psi(n,m+1) + T_m \Psi(n,m-1)
  = E\Psi(n,m),
\end{equation}
where $\Psi(n,m)$ is the value of a $2d$ eigenfunction on a vertex
labeled by the two integers $n$ and $m$, and $E$ is the corresponding
eigenvalue. The hopping amplitudes $T_j$ are equal to 1 or $T$
according to the Fibonacci sequence
$\{T_j\}=\{1,T,1,1,T,1,T,1,1,T,1,1,T,1,T,1,1,T,1,T,\ldots\}$. By prohibiting
diagonal hopping, the resulting high-dimensional eigenvalue problem is
ensured to be separable. This allows one to use the known solutions
for the $1d$
problem~\cite{1dmodel1,1dmodel2,1dmodel3,1dmodel4,review1,review2,%
  review3,review4} in order to construct the solutions in two and
higher dimensions (as was done for similar models in the
past~\cite{2dmodel1,2dmodel2,2dmodel3,2dmodel4,2dmodel5,2dmodel6,ashraff}).
Two-dimensional eigenfunctions can therefore be expressed as Cartesian
products of the $1d$ eigenfunctions~\cite{me}, and the corresponding
$2d$ eigenvalues are given by pairwise sums of the $1d$ eigenvalues.

The $1d$ spectrum for the $N^{th}$ order Fibonacci approximant is
composed of $F_N$ bands, where $F_N = F_{N-1} + F_{N-2}$ is the
$N^{th}$ Fibonacci number, starting with $F_0=F_1=1$. The edges of
each such band correspond to either periodic or antiperiodic boundary
conditions. Hence, by direct diagonalization of the two corresponding
hamiltonians for a single approximant we obtain the edges of the
energy intervals in the spectrum. The $2d$ and $3d$ spectra are then
calculated as the Minkowski sums of two or three $1d$ spectra, where
the Minkowski sum of two sets $A$ and $B$ is the result of adding
every element of $A$ to every element of $B$, {\it i.e.}  the set
\begin{equation}\label{eq:defsum}
A+B=\left\{x+y\ \vert\ x\in A,\ y\in B\right\}.
\end{equation}

Although the spectrum of the $1d$ Fibonacci model, for any choice of
$T\neq1$, is a totally disconnected set with zero bandwidth and an
infinite number of bands, the higher-dimensional spectra exhibit
different behavior for different values of the relative hopping
parameter $T$, including spectra that contain continuous intervals and
have a finite measure~\cite{ilan}. A similar situation arises in the
case of the well-known {\it ternary Cantor set}~\cite{cantor}, which
is constructed iteratively by starting with the closed interval
$[0,1]$, and at each iteration removing the open middle thirds of all
remaining closed intervals from the previous iteration. The first few
approximants that are obtained in this way are $C_0=[0,1]$,
$C_1=[0,1/3] \cup [2/3,1]$, and $C_2=[0,1/9] \cup [2/9,1/3] \cup
[2/3,7/9] \cup [8/9,1]$, so that after $N$ such iterations one is left
with an approximant set $C_N$ consisting of $2^N$ closed intervals,
each of which has a measure $1/3^N$, and therefore the total measure
of the set is $\left(2/3\right)^N$.  The ternary Cantor set itself
$C_\infty$, defined as the limit $N\rightarrow\infty$ of this sequence
of sets, contains uncountably-many points yet no interval, it is
totally disconnected, and its total measure is zero. By simple
inspection, one finds that for any finite order Cantor approximant
$C_N$, the Minkowski sum $C_N+C_N$ is the entire interval $[0,2]$. One
can show that this also holds in the limit $N\to\infty$, namely that
$C_\infty+C_\infty=[0,2]$. Thus, even though $C_\infty$ contains no
interval, its sum with itself covers the whole interval from 0 to 2.

For a given dimension $n$, we identify a sequence of values $1<
\Tn1\leq \Tn2\leq \Tn3\leq \Tn4$ corresponding to the following
transitions in the spectrum: 
\begin{enumerate}
\item[$\Tn1$:] The value of $T$ below which all bands in the
  $n$-dimensional spectrum are of positive, finite measure. For
  $T>\T{n}1$ there is at least a finite number of zero measure bands in
  the spectrum.
  
\item[$\Tn2$:] The value of $T$ above which the number of bands in the
  $n$-dimensional spectrum is infinite. An infinite number of bands in
  a spectrum of finite bandwidth necessarily implies that infinitely
  many bands are of zero measure, thus $\Tn2\geq \Tn1$.
  
\item[$\Tn3$:] The value of $T$ above which all bands in the spectrum
  are of zero measure.\footnote{Note that the absence of intervals in
    the spectrum above $\Tn3$ does not necessarily correspond to zero
    total bandwidth. It is in fact possible to use the Cantor set
    generation process to obtain a totally disconnected set with a
    finite measure. For example, if at the $N^{th}$ iteration of the
    generation process the middle $1/3^N$ part is removed from
    each of the remaining intervals, one ends with a totally
    disconnected set whose measure is $\lim
    \prod_{k=1}^{\infty}\left(1-{1}/{3^k}\right)\simeq 0.5601.$}

\item[$\Tn4$:] The value of $T$ above which the total bandwidth of the
  spectrum is zero.
\end{enumerate}

We use two different approaches to study the behavior of the spectrum.
In Sec.~\ref{sec:cantor} we use analytical results derived for the
addition of generalized Cantor sets to obtain an upper bound on the
transition between absolutely continuous and singular continuous
spectra. In Sec.~\ref{sec:direct} we use direct numerical calculation
of the $2d$ and $3d$ spectra of Fibonacci approximants of finite order
to extrapolate for the behavior in the quasiperiodic limit. In an
earlier paper\cite{me} we studied only two of the transitions,
$\Tn2$ and $\Tn4$. To find $\Tn4$ we used a naive method based
on the results of Ashraff {\it et al.}~\cite{ashraff} for the diagonal
tight-binding hamiltonian. The current results include a correction to
our previous calculation. In Sec.~\ref{sec:dynamics} we summarize the
results, and discuss their expected relation to the nature of
eigenfunctions and to quantum dynamics, indicating directions for
future work.

\section{Analytical bounds for the appearance of continuous intervals
  in the spectrum}
\label{sec:cantor}

\subsection{Addition of generalized Cantor sets - Known results}
\label{sec:generalized}

A generalized Cantor set is obtained just like the ternary Cantor set
except that the open intervals removed at each iteration are not
necessarily the middle thirds of the remaining closed intervals. For
each interval removed from the set, one defines a {\it left (right)
  ratio of dissection\/} as the ratio between the length of the left
(right) remaining interval and the length of the original one. Sets
for which the left and right ratios are the same are called
\textit{central Cantor sets}.  In general, the ratios of dissection
may vary between the left and right resulting intervals, between
different iterations of the process, and between different intervals
at the same step. The ternary Cantor set is a central Cantor set with
a constant ratio of dissection of $1/3$.

We are interested in conditions
for the appearance of intervals in the Minkowski sum of $n$
generalized Cantor sets. For central Cantor sets with a constant ratio
of dissection $a$, one can show that the condition for the sum to be an
interval is
\begin{equation}\label{eq:cabrelli2}
n\frac{a}{1-a}\geq 1\qquad {\rm or\ }\qquad a\geq\frac1{n+1}.
\end{equation}
Thus, the ternary Cantor set exactly has the critical value of $a=1/3$
for which a sum of $n=2$ central Cantor sets is an interval. Cabrelli
{\it et al.}~\cite{Cabrelli} found, more generally, a sufficient
condition for the existence of an interval in the sum of $n$
generalized Cantor sets, all of which can be constructed with a lower
bound $a$ on their ratios of dissection, which is given by
\begin{equation}\label{eq:cabrelli}
(n-1)\frac{a^2}{(1-a)^3}+\frac{a}{1-a}\geq 1.
\end{equation}

\subsection{Applying Cantor set results to the Fibonacci spectra}
\label{sec:applying}

Before using the results quoted above to analyze the Fibonacci
spectra, we should note that there exist two important differences
between the energy spectra $S_N$ of the $N^{th}$ order approximants of
the $1d$ Fibonacci quasicrystal, and finite approximants $C_N$ of
generalized Cantor sets. The spectrum $S_N$ consists of $F_N$ rather
than $2^N$ energy intervals, and is not contained in the spectrum
$S_{N-1}$ of the approximant of order $N-1$. One should therefore take
care in defining the spectrum $S$ of the Fibonacci quasicrystal itself
as the set of limit values for sequences of energies taken from
consecutive spectra $S_N$ of finite order approximants
\begin{equation}\label{eq:defspectrum}
S=\left\{E=\lim_{n\rightarrow\infty}E_n\ \vert\ E_n\in S_n\right\}.
\end{equation}

The fact that the number of bands in $S_N$ is $F_N$ rather than $2^N$
implies that the spectra cannot be constructed by the iterative
process described above for generalized Cantor sets, and hence that
the ratios of dissection cannot be defined. However, the spectrum
$S_N$ of a finite approximant can be padded with additional intervals
which can be chosen in a manner that will not disturb the calculation,
and will increase the number of intervals to $2^N$, as in the Cantor
approximant.  This allows to calculate backwards and define
\textit{effective ratios of dissection}. The additional intervals can
be added on either, or both, ends of the spectrum. Thus, the effective
ratios of dissection are not uniquely determined.

\begin{figure}[h]
\hspace*{0pt}
\begin{center}
  \scalebox{0.50}{\rotatebox{00}{\includegraphics{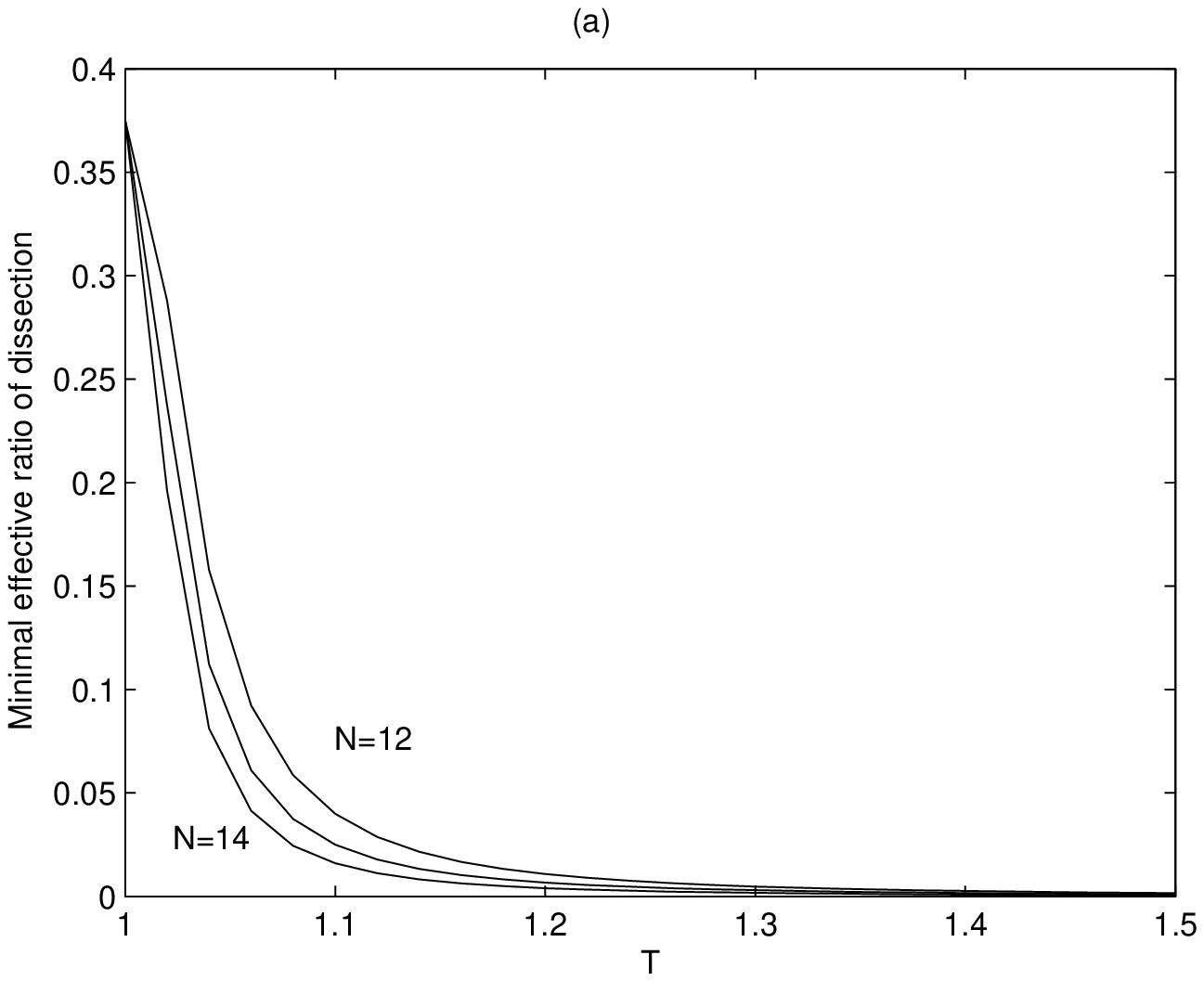}}}
  \scalebox{0.50}{\rotatebox{00}{\includegraphics{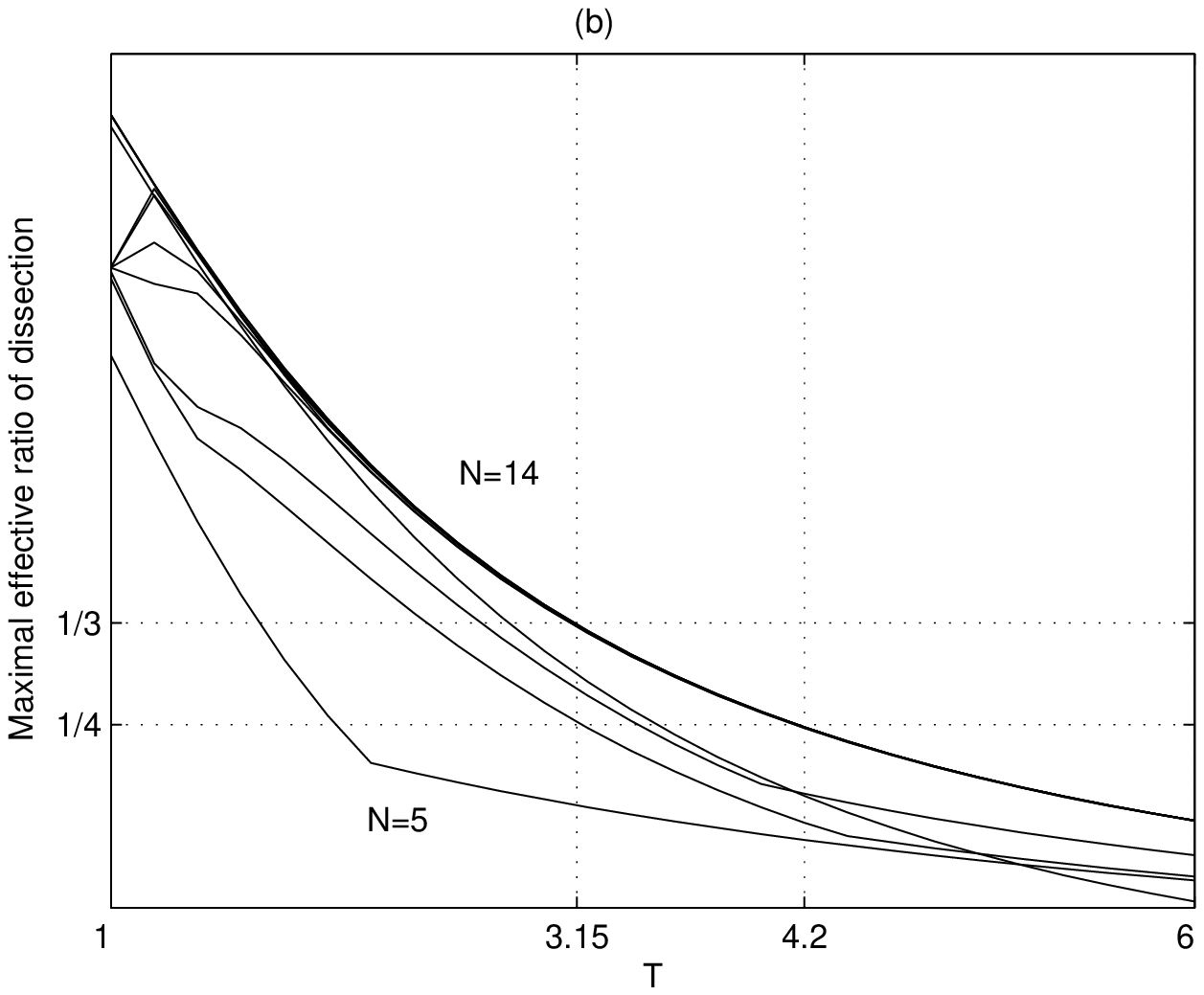}}}
\end{center}
\caption{(a) The minimal effective ratio of dissection calculated
  for the $1d$ spectra of approximants of order 12-14 for values of
  $T$ up to 1.5. The sharp drop near $T=1$ means that no value of $T$
  satisfies the sufficient condition for obtaining an interval in the
  higher-dimensional spectra. (b) The maximal effective ratio of
  dissection calculated for the $1d$ spectra of approximants of order
  5-14 for values of $T$ up to 6. The horizontal dotted lines are
  drawn at $1/3$ and $1/4$ to indicate the upper bounds for the value
  of $T$ at which no intervals are to appear in the $2d$ and $3d$
  spectra respectively.
\label{fig:minmaxratio}}
\end{figure}

We have tried using Eq.~(\ref{eq:cabrelli}) to find a sufficient
condition for the higher-dimensional spectra to contain an interval.
This would provide a lower bound on $\Tn3$---a value of $T$ below
which the condition is satisfied and the $n$-dimensional spectrum
necessarily contains an interval.  Unfortunately, as one studies the
effective ratios of dissection defined for the $1d$ spectrum it turns
out that regardless of the way in which the approximant spectra are
embedded in Cantor approximants, the ratios of dissection are not
bounded away from zero, even for small values of $T$, as shown in
Fig.~\ref{fig:minmaxratio}(a). Hence, at this point we do not know how
to use the condition of Cabrelli {\it et al.} to obtain a lower bound
on $\Tn3$.

Nevertheless, by studying the maximal effective ratio of dissection we
can obtain an upper bound for the value of $\Tn3$ above which the
higher-dimensional spectra do not contain an interval.
Fig.~\ref{fig:minmaxratio}(b) shows the effective maximal ratio for
approximants of order $N=5$ ($F_N=8$) to $N=14$ ($F_N=610$).  It is
evident that the maximal ratio of dissection rapidly converges as a
function of the order of the approximant, with almost no difference
between the the curves for $N=9$ and above.  It is also of interest to
note that the maximal ratio is independent of the way in which the
approximant spectrum is embedded in a Cantor approximant. Values of
$T$ for which the maximal ratio of dissection fails to satisfy
Eq.~(\ref{eq:cabrelli2}) imply that there is no portion of the $1d$
spectrum which can lead to the existence of an interval in the
higher-dimensional spectra. The maximal ratio of dissection becomes
$1/3$ at $T\simeq 3.15$ and $1/4$ at $T\simeq 4.2$. Thus, we expect to
see the vanishing of intervals in the spectrum at a value of $T$ below
these upper bounds for $2d$ and $3d$ respectively.

\section{Direct study of the $2d$ and $3d$ spectra}
\label{sec:direct}

We now turn to the direct study of the higher-dimensional spectra.
This is done by explicitly calculating the spectra for approximants of
finite order. Each pair or triplet of energy bands in the $1d$
spectrum is summed to yield a single band in the $2d$ or $3d$
spectrum, respectively. A set of $F_N$ bands in the $1d$ spectrum
generates ${(F_N+1)F_N}/{2}$ bands in the corresponding $2d$ spectrum,
and ${(F_N+2)(F_N+1)F_N}/{6}$ bands in the $3d$ spectrum, with
possible overlaps that decrease as $T$ increases. Overlapping bands
are merged into single energy intervals to obtain the actual structure
of the higher-dimensional spectra. Note that we shall use the term
`bands' to refer to the continuous energy intervals in the spectra,
even though strictly speaking they may be composed of different bands
with overlapping energies.

\subsection{Measuring the smallest band in the spectrum to find $\Tn1$}
\label{sec:smallest}

\begin{figure}[tb]
\hspace*{0pt}
\begin{center}
  \scalebox{0.50}{\rotatebox{00}{\includegraphics*{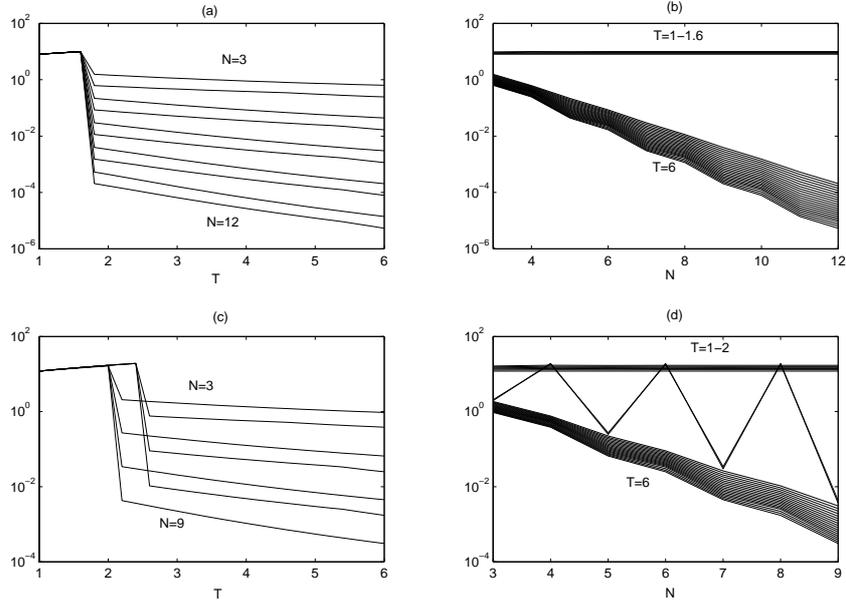}}}
  \caption{The length of the smallest band $B_{min}$ in the spectrum of the
    $2d$ (top) and the $3d$ (bottom) Fibonacci quasicrystals.  The
    length of the smallest band is plotted on the left as a function
    of $T$ for different approximants, and on the right as a function
    of $N$ for different values of $T$. The linear slopes in the
    semi-logarithmic plots as a function of $N$ indicate a power law
    behavior, $B_{min}\propto \tau^{-N\alpha_n(T)}$. The exponents
    $\alpha_n(T)$ are plotted in Fig.~\ref{fig:smallestexponent}.
\label{fig:smallest}}
\end{center}
\end{figure}

For $T>\Tn1$ there is at least one zero-measure band in the
spectrum.  We therefore measure the smallest band $B_{min}$ and ask
whether it vanishes in the limit of $N\to\infty$. For $T<\Tn1$ the
length of the smallest band is independent of the order $N$ of the
approximant.  For $T>\Tn1$ it can be described by a power law
$B_{min}\propto F_N^{-\alpha_n(T)}$ with some positive exponent,
$\alpha_n(T)$. We locate $\T{n}1$ by finding the value of $T$ for
which $\alpha_n$ vanishes.  Fig.~\ref{fig:smallest}(b) clearly shows
that $1.6<\T21<1.8$, and Fig.~\ref{fig:smallest}(d) indicates that
$2<\T31<2.6$. Within these bounds, the width of the smallest band
oscillates between the two different limiting behaviors.

\begin{figure}[tb]
\hspace*{0pt}
\begin{center}
  \scalebox{0.45}{\rotatebox{00}{\includegraphics*{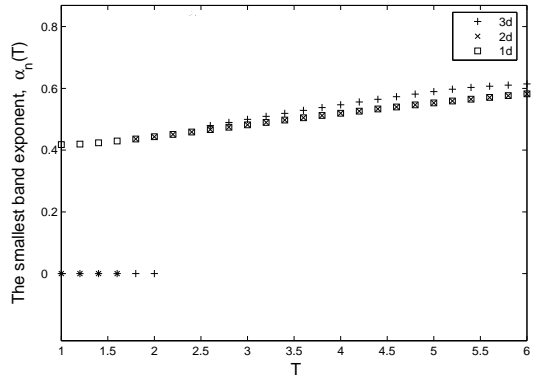}}}
  \caption{The exponents $\alpha_n(T)$ extracted from
    Fig.~\ref{fig:smallest}. The curves for $1d$, $2d$, and $3d$ all
    coincide for values of $T$ above the transition at $\Tn1$.
\label{fig:smallestexponent}}
\end{center}
\end{figure}

As $T$ increases and the overlap of bands vanishes, the smallest band
in the $n$-dimensional spectrum is expected to be $n$ times the
smallest band of the $1d$ spectrum. Hence for high values of $T$ the
exponents $\alpha_n(T)$ should be independent of the dimension,
because the multiplicative factor of $n$ only adds a constant term in
the semi-logarithmic scale. Fig.~\ref{fig:smallestexponent} shows the
extracted exponents $\alpha_n(T)$, indicating that they indeed
coincide for all values of $T$ above $\Tn1$.

\subsection{Counting the number of bands to find $\Tn2$}
\label{sec:number}

Next we count the number of bands $\#B$ in the spectrum and ask
whether it tends to infinity or remains finite as $N$ increases.
Again, we express this number as a power law of the form $\#B\propto
F_N^{\beta_n(T)}$, expecting $\beta_n(T)$ to vanish for $T<\Tn2$. For
the $1d$ Fibonacci quasicrystal $\#B_1 = F_N \propto \tau^N$, where
$\tau$ is the golden mean.  In higher dimensions, as the overlap
between bands decreases with increasing $T$, we expect the number of
bands to tend to its maximal value, which is approximately
${(\#B_{1})^2}/{2}$ in $2d$, and approximately ${(\#B_{1})^3}/{6}$ in
$3d$.  Thus the exponents $\beta_n(T)$ should tend to $n\log\tau$ as
$T\rightarrow\infty$. The dashed horizontal line in
Fig.~\ref{fig:numberexponent} indicates the expected limit value for
the $2d$ model which indeed tends to it. For the $3d$ model the limit
is only obtained at significantly higher values of $T$, indicating
that the overlap of bands plays a significant role in the structure of
the spectrum even at relatively high values of $T$. The continuous
variation of $\beta_2(T)$ allows us to use smooth extrapolation and
find $\T22\simeq 1.66$, whereas in $3d$ we can only conclude that $2.0
< \T32 < 2.6$. Combining the fact that $\Tn2\geq \Tn1$ with the
results for the exponents $\beta_n(T)$ as shown in
Fig.~\ref{fig:numberexponent}, we find that at least in $2d$ and $3d$,
$\Tn2 = \Tn1$, and hence that there is no intermediate regime in which
the spectrum contains only a finite number of zero-measure bands.

\subsection{Measuring the largest band in the spectrum to find $\Tn3$}
\label{sec:largest}

\begin{figure}[tb]
\hspace*{0pt}
\begin{center}
  \scalebox{0.50}{\rotatebox{00}{\includegraphics*{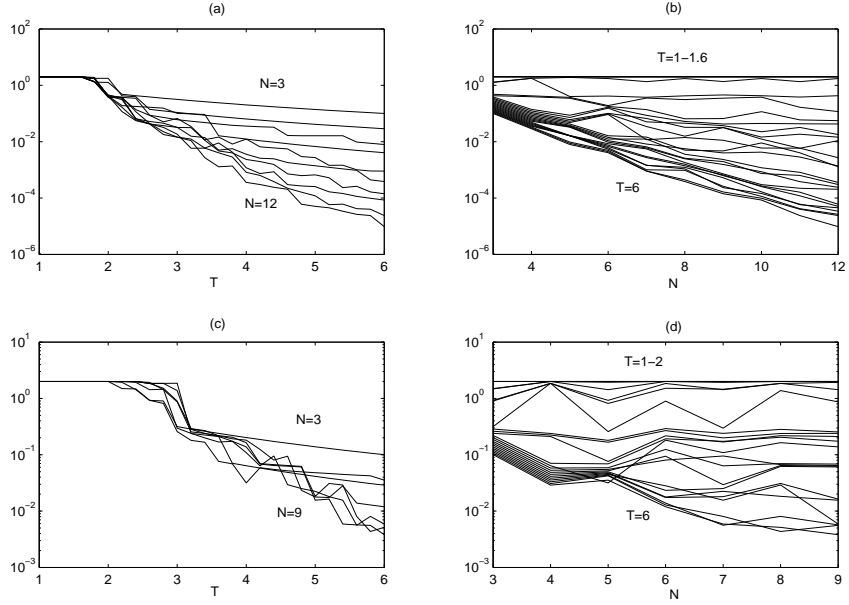}}}
  \caption{The normalized length $B_{max}$ of the largest band in the
    spectrum of the $2d$ (top) and the $3d$ (bottom) Fibonacci
    quasicrystals.  $B_{max}$ is plotted on the left as a
    function of $T$ for different approximants, and on the right as a
    function of $N$ for different values of $T$. The linear slopes in
    the semi-logarithmic plots as a function of $N$ indicate a power
    law behavior, $B_{max}\propto \tau^{-N\gamma_n(T)}$. The exponent
    $\gamma_2(T)$ is plotted in Fig.~\ref{fig:largestexponent}.
\label{fig:largest}}
\end{center}
\end{figure}

\begin{figure}[tb]
\hspace*{0pt}
\begin{center}
  \scalebox{0.45}{\rotatebox{00}{\includegraphics*{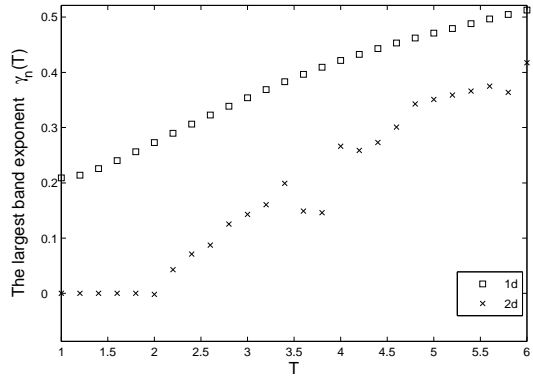}}}
\end{center}
\caption{The exponent $\gamma_2(T)$ extracted from
  Fig.~\ref{fig:largest} and compared with $\gamma_1(T)$. The
  asymptotic behavior in which all three curves are expected to
  coincide is not observed for the values of $T$ shown.
\label{fig:largestexponent}}
\end{figure}

For $T>\Tn3$ all bands in the spectrum have zero measure. We therefore
look at the width of the largest band in the spectrum and ask whether
it vanishes as $N\to\infty$.  However, since the maximal energy in the
spectrum is approximately $n(1+T)$, for small values of $T$ the
overlap of bands leads to an increase in the width of largest band as
a function of $T$.  To avoid this we normalize the results, dividing
by the maximal energy in the spectrum. Thus, for $T>\Tn3$, we express
the normalized largest band as a power law $B_{max}\propto
F_n^{-\gamma_n(T)}$. Figs.~\ref{fig:largest}(b) and
\ref{fig:largestexponent} clearly indicate that $\T23\simeq 2$, but in
$3d$ oscillatory behavior dominates a large range of values for $T$,
and we cannot determine the transition without extending the analysis
to higher order approximants. However, from Fig.~\ref{fig:largest}(d)
we can infer that the transition occurs at some value of $T$ below 5,
for which we obtained analytically a stricter upper bound of $\T33\leq
4.2$ as shown in Fig.~\ref{fig:minmaxratio}(b).

As for $B_{min}$, at large values of $T$, $B_{max}$ is also
expected to be $n$ times the largest band of the $1d$ spectrum, and
hence the exponents should be independent of dimension. The fact that
this does not occur indicates, once again, that the overlap of bands
is still significant for values of $T$ as large as 6.

\subsection{Calculating the total measure of the spectrum to find
  $\Tn4$}
\label{sec:total}

\begin{figure}[tb]
\hspace*{0pt}
\begin{center}
  \scalebox{0.50}{\rotatebox{00}{\includegraphics*{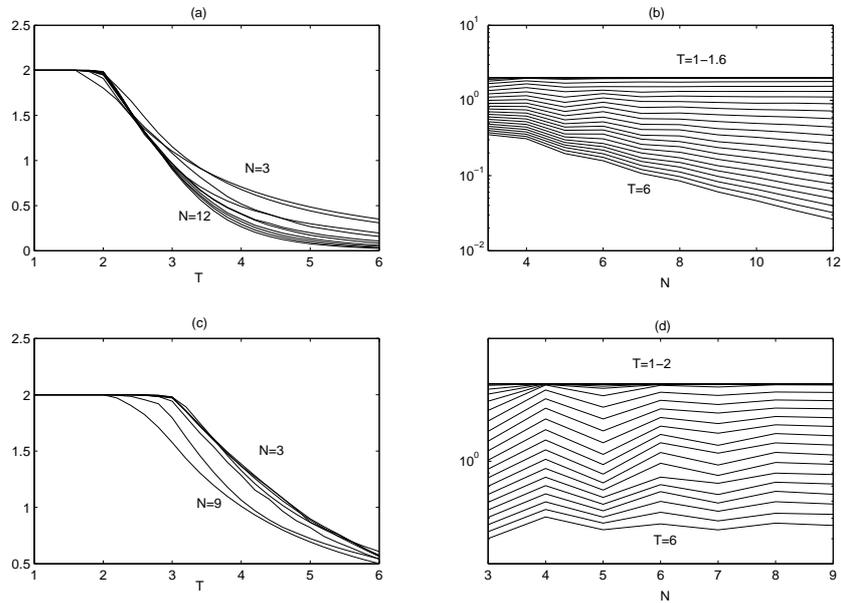}}}
\end{center}
\caption{The normalized total bandwidth $W$ of the spectrum of the
  $2d$ (top) and the $3d$ (bottom) Fibonacci quasicrystals.  $W$ is
  plotted on the left as a function of $T$ for different approximants,
  and on the right as a function of $N$ for different values of $T$.
  The linear slopes in the semi-logarithmic plots as a function of $N$
  indicate a power law behavior, $W\propto \tau^{-N\delta_n(T)}$. The
  exponents $\delta_n(T)$ are plotted in Fig.~\ref{fig:totalexponent}.
\label{fig:total}}
\end{figure}

\begin{figure}[tb]
\hspace*{0pt}
\begin{center}
  \scalebox{0.45}{\rotatebox{00}{\includegraphics*{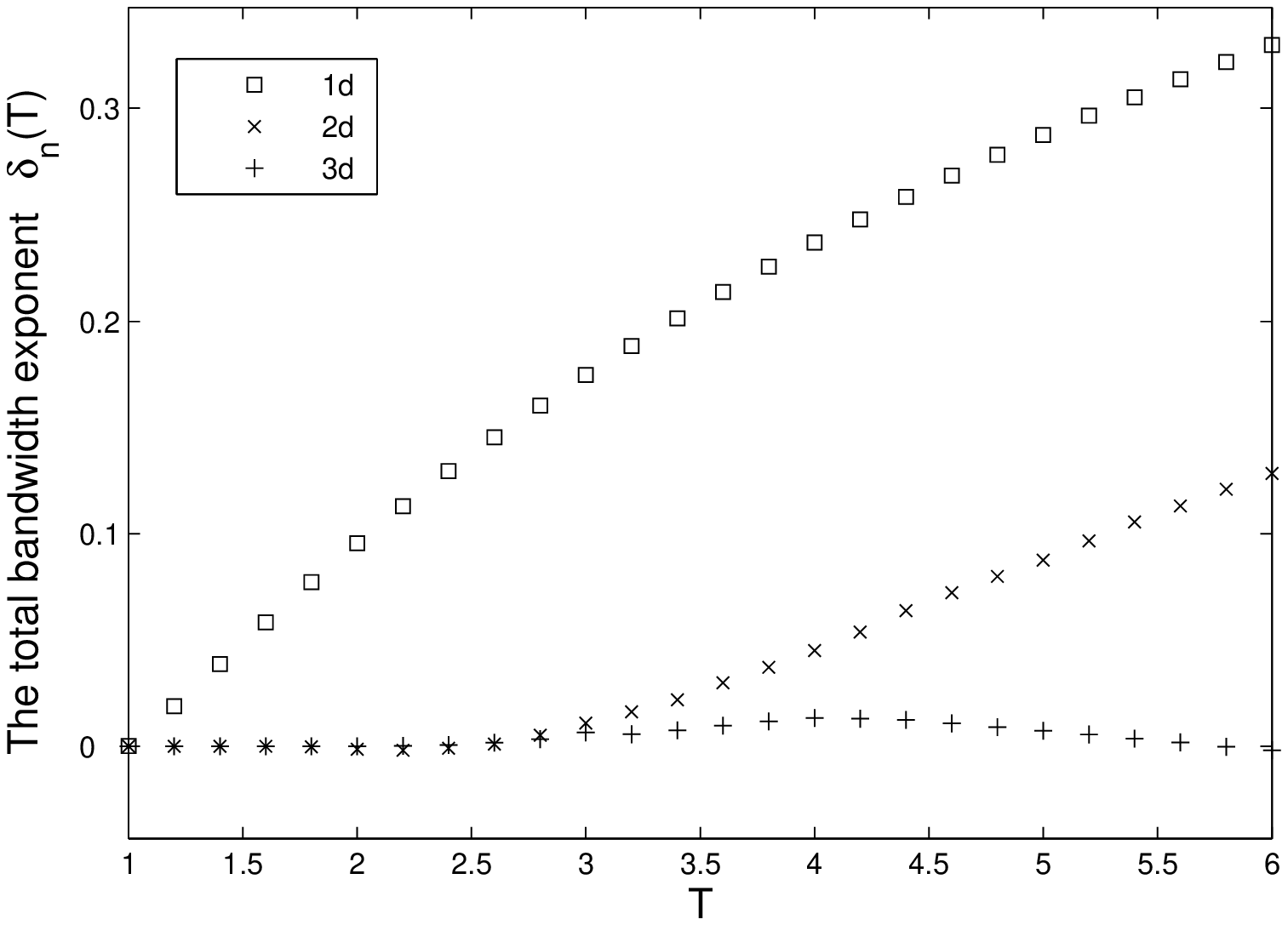}}}
  \caption{The exponents $\delta_n(T)$, extracted from
    Fig.~\ref{fig:total}. In $3d$ it is evident that the transition to
    zero bandwidth does not occur within the studied range of $T$
    values.
\label{fig:totalexponent}}
\end{center}
\end{figure}

To find $\Tn4$ we measure the total bandwidths of the spectra as $N$
increases, normalizing by $1+T$, and looking for a power law decay of
the normalized bandwidth $W\propto F_N^{-\delta_n(T)}$.
Fig.~\ref{fig:total}(c) shows a decrease in the normalized total
measure of the spectrum as a function of $T$ in $2d$, but
Fig.~\ref{fig:total}(d) shows the total measure in $3d$ to be almost
independent of $N$ for any given value of $T$. Thus, although the $3d$
spectrum consists only of zero measure bands for values of $T$ above
5, its total measure remains finite over the entire range of $T$
values studied. The exponents $\delta_n(T)$ are shown in
Fig.~\ref{fig:totalexponent}. The transition to zero total bandwidth
in $2d$ occurs at $\T24\simeq 2.6$. In $3d$ we can only say that
$\T34>6$.

\section{Summary and future work}
\label{sec:dynamics}

The results of Sections~\ref{sec:cantor} and~\ref{sec:direct} are
summarized as follows
\begin{equation}
\begin{tabular}{|c|c|c|c|c|}
  \hline
   & $\Tn1=\Tn2$ & $\Tn3$ & Upper bound for $\Tn3$& $\Tn4$ \\
  \hline
  $2d$ & $\sim 1.66$ & $\sim 2$ & 3.15 & $\sim 2.6$ \\\hline
  $3d$ & $2.0 - 2.6$ & $\leq5$ & 4.2 & $>6$ \\\hline
\end{tabular}
\end{equation}
The transitions between different regimes in the spectrum are expected
to reflect on the physical properties of the Fibonacci quasicrystals,
on the nature of eigenfunctions and on the dynamics of electronic wave
packets. For values of $T$ above the transition $\Tn3$ the
higher-dimensional spectra are similar to the $1d$ spectrum in being
totally disconnected, singular continuous sets, and hence the
eigenfunctions are expected to be critical, and wave packets are
expected to display sub-ballistic dynamics. Note that the last
transition $\Tn4$ is of no consequence for this matter because the
spectrum is purely singular continuous both above and below this value.
For values of $T$ below the lowest transition point $\Tn1=\Tn2$, where
the spectra are absolutely continuous we expect to find extended
eigenfunctions, and wave packets are expected to display ballistic
dynamics. For the intermediate range between these transitions the
spectra contain both singular continuous and absolutely continuous
parts, and therefore we expect to find mixed ballistic and
sub-ballistic dynamics, and some of the wave functions to be extended.

We intend to complement these studies by simulating the dynamics of
electronic wave functions to find whether transition between ballistic
and sub-ballistic dynamics occur at the points found here. We also
intend to use the degeneracy of wave functions in the $2d$ Fibonacci
quasicrystal (as hypothesized in Ref.~\cite{ilan}) to construct
maximally extended wave functions, again, we expect to find some
qualitative change in the nature of these wave functions near the
transition points indicated above.

\section*{Acknowledgments}

This research is supported by the Israel Science Foundation
through Grant No.~684/06.

\singlespacing

%##################### Figure 1####################################
%##################### Figure 1####################################

\newpage
%##################### Figure 2####################################
%##################### Figure 2####################################

%##################### Figure 3####################################
%##################### Figure 3####################################

%##################### Figure 4####################################
\begin{figure}[tb]
\hspace*{0pt}
\begin{center}
  \scalebox{0.50}{\rotatebox{00}{\includegraphics*{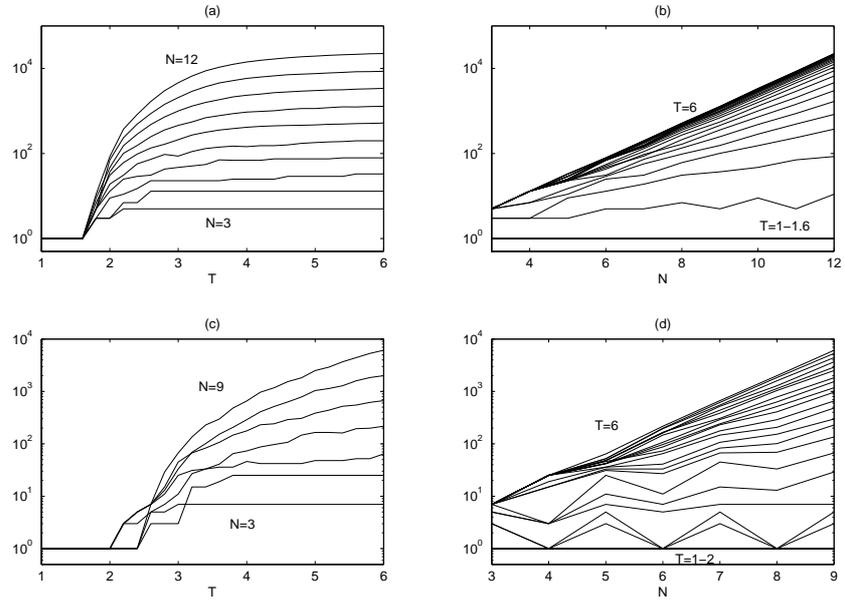}}}
  \caption{The number of bands $\#B$ in the spectrum of the
    $2d$ (top) and the $3d$ (bottom) Fibonacci quasicrystals.  The
    number of bands is plotted on the left as a function
    of $T$ for different approximants, and on the right as a function
    of $N$ for different values of $T$. The linear slopes in the
    semi-logarithmic plots as a function of $N$ indicate a power law
    behavior, $\#B\propto \tau^{N\beta_n(T)}$. The exponents
    $\beta_n(T)$ are plotted in Fig.~\ref{fig:numberexponent}.
\label{fig:number}}
\end{center}
\end{figure}
%##################### Figure 4####################################

%##################### Figure 5####################################
\begin{figure}[tb]
\hspace*{0pt}
\begin{center}
  \scalebox{0.45}{\rotatebox{00}{\includegraphics*{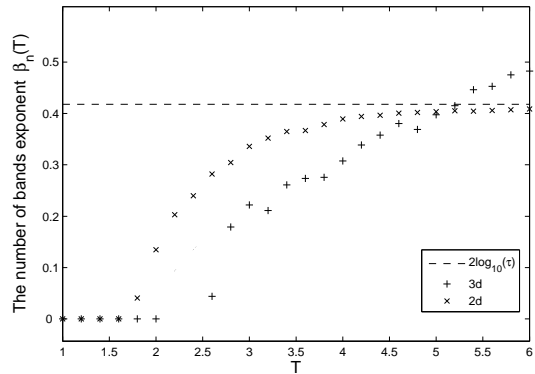}}}
  \caption{The exponents $\beta_n(T)$, extracted from
    Fig.~\ref{fig:number}. The horizontal dashed line indicates the
    expected asymptotic value of $2\log\tau\simeq 0.418$ for the $2d$
    quasicrystal.
\label{fig:numberexponent}}
\end{center}
\end{figure}
%##################### Figure 5####################################

%##################### Figure 6####################################
%##################### Figure 6####################################

%##################### Figure 7####################################
%##################### Figure 7####################################

%##################### Figure 8####################################
%##################### Figure 8####################################

%##################### Figure 9####################################
%##################### Figure 9####################################


\begin{thebibliography}{99}
  
\bibitem{shechtman} D.~Shechtman, I.~Blech, D. Gratias, and J.W.~Cahn,
  Phys. Rev. Lett. {\bf 53}, 1951 (1984).

\bibitem{ilan} R.~Ilan, E.~Liberty, S.~Even-Dar~Mandel, and
R.~Lifshitz, Ferroelectrics {\bf 305}, 15 (2004).

\bibitem{me} S.~Even-Dar Mandel and R.~Lifshitz, Phil. Mag. {\bf 86},
  759 (2006). 

\bibitem{squarefib} R.~Lifshitz. J. of Alloys and Compounds {\bf 342},
  186 (2002).

\bibitem{definition1} R.~Lifshitz. Z. Kristallogr. {\bf 222}, 313 (2007).

\bibitem{definition2} R.~Lifshitz. Foundations of Physics {\bf 33},
  1703 (2003). 

\bibitem{1dmodel1} M.~Kohmoto, L.P.~Kadanoff, and C.~Tang, Phys. Rev.
  Lett. {\bf 50}, 1870 (1983).
  
\bibitem{1dmodel2} S.~Ostlund, R.~Pandit, D.~Rand, H.S.~Schellnhuber,
  and E.D.~Siggia, Phys. Rev. Lett. {\bf 50}, 1873 (1983).
  
\bibitem{1dmodel3} M.~Kohmoto and J.R.~Banavar, Phys. Rev. B {\bf 34},
  563 (1986).
  
\bibitem{1dmodel4} M.~Kohmoto, B.~Sutherland, and C.~Tang, Phys. Rev.
  B {\bf 35}, 1020 (1987).
  
\bibitem{review1} T.~Janssen, in {\it The Mathematics of Long-Range
    Aperiodic Order}, ed. R.V.~Moody, (Kluwer, Dordrecht, 1997) p.~269.
    
\bibitem{review2} T.~Fujiwara, in {\it Physical Properties of
    Quasicrystals}, ed. Z.M.~Stadnik, (Springer, Berlin, 1999) ch.~6.
    
\bibitem{review3} J.~Hafner and M.~Kraj{c}\'{\i}, {\it ibid.} ch.~7.
    
\bibitem{review4} D.~Damanik, in {\it Directions in Mathematical
    Quasicrystals}, ed. M.~Baake and R.V.~Moody, (AMS, Providence,
    2000) p.~277.

\bibitem{2dmodel1} K.~Ueda and H.~Tsunetsugu, Phys. Rev. Lett. {\bf 58},
  1272 (1987).

\bibitem{2dmodel2} W.A.~Schwalm and M.K.~Schwalm, Phys. Rev. B {\bf 37},
  9524 (1988).
  
\bibitem{2dmodel3} J.X.~Zhong and R.~Mosseri, J. Phys: Condens. Matter
  {\bf 7}, 8383 (1995).
  
\bibitem{2dmodel4} S.~Roche and D.~Mayou, Phys. Rev. Lett. {\bf 79},
  2518 (1997).
  
\bibitem{2dmodel5} Yu.Kh.~Vekilov, I.A.~~Gordeev, and E.I.~Isaev, JETP
  {\bf 89}, 995 (1999).
  
  
\bibitem{2dmodel6} Yu.Kh.~Vekilov, E.I.~Isaev, I.A.~Gordeev, Mat. Sci.
  and Eng. {\bf 294-296}, 553 (2000).

\bibitem{ashraff}  J.A.~Ashraff, J.-M.~Luck, and R.B.~Stinchcombe,
  Phys. Rev. B {\bf 41}, 4314 (1990).
  
\bibitem{cantor} G.~Cantor, ``De la puissance des ensembles parfait de
  points'' (On the Power of Perfect Sets of Points), Acta Mathematica
  {\bf 4}, 381 (1884). English translation reprinted in {\it Classics on
  Fractals}, ed. Gerald A. Edgar, (Addison-Wesley, 1993).

\bibitem{Cabrelli} C.A.~Cabrelli, K.E.~Hare and U.M.~Molter J. Aust.
  Math. Soc. {\bf 73}, 405 (2002).


\end{thebibliography}
\end{document}